\documentclass[preprint,showpacs,preprintnumbers,amsmath,amssymb]{revtex4}

\usepackage{graphicx}
\usepackage{dcolumn}
\usepackage{bm}

\begin{document}

\preprint{APS/123-QED}

\title{Correlation between balance energy and transition energy for symmetric
colliding nuclei.\\}
\author{Rajni}
\author{Suneel Kumar}%
 \email{suneel.kumar@thapar.edu}
\affiliation{%
School of Physics and Materials Science, Thapar University, Patiala-147004, Punjab (India)\\
}%

\author{Rajeev K. Puri}
\affiliation{
Department of Physics, Panjab University, Chandigarh-160014 (India)\\
}%

\date{\today}

\begin{abstract}
We study the  correlation between balance energy and transition energy of 
fragment in heavy-ion collisions for different systems at incident energies
between 40 and 1200 MeV/nucleon using an isospin-dependent quantum molecular dynamics model. With increasing
incident energy, the elliptic flow shows a transition from positive (in-plane) to negative (out-of-plane)
flow. This transition energy is found to depend on the size of fragments, composite mass of reacting system, and
the impact parameter of reaction. It has been observed that reduced cross-section can explain
the experimental data. There is a correlation between transition energy and balance energy
as their difference decreases with increase in the total mass of colliding nuclei.
\end{abstract}


\maketitle
\section{Introduction}
The elliptic flow is a measure that quantifies the azimuthal anisotropy of the momentum distribution.
Specifically, we fit the azimuthal distribution of nucleons about reaction plane with a Fourier expansion of the form:
\begin{equation} \frac{dN}{d\phi} = v_0(1 + 2v_1Cos\phi + 2v_2Cos2\phi + .....).
\end{equation}
where, $v_0$ is for normalization only, $v_1$ characterizes the directed in-plane flow, while $v_2 > 0$ indicates
in-plane enhancement, $v_2 < 0$ characterizes the squeeze-out perpendicular to the reaction plane, and  $v_2 = 0$
shows an isotropic distribution of nucleon momentum in the transverse plane.
Hence, the ellipticity coefficient $v_2$ depends on the in-plane and 
out-of-plane flow amplitudes.
        The elliptical flow parameters $\langle cos2\phi \rangle$ at energies 
from tens to hundreds of MeV per nucleon are determined by the complex 
interplay among expansion, rotation, and the shadowing of spectators.
Both the mean field and two body collision parts play important roles 
in this energy region. The mean field play a dominant role at low energies, 
and then gradually the two body collision becomes dominant
with increase in energy. Two colliding nuclei create a stopped overlap 
region. At higher bombarding energies ($E_{lab}\ge
1 GeV/nucleon$) the spectator leave interaction zone rapidly. The remaining interaction zone expands almost
freely, where the surface is such that in-plane emission is preferred. It is therefore also the interplay
between the timescales of passing time of spectators and expansion time of the dense, stopped interaction zone
which detemines the time-integrated elliptic flow signal. Experimentally observed out-of-plane emission, termed as
squeeze was first observed by at SATURNE (France) by the DIOGENE Collaboration \cite{Gosset77}.
Plastic Ball group at the BEVALAC in BERKLEY were the first one to quantify the
squeeze out in symmetric systems \cite{Renfo84}.
  Recently, elliptic flow has been measured at relativistic heavy 
ion collider (RHIC) in Au+Au collision at
$\sqrt s = 130 GeV/nucleon$ \cite{lin02}. At top AGS and SPS energies, 
elliptic flow is inferred to be a relative
enhancement of emission in the plane of the reaction. Elliptic flow is 
developed mostly in the first several fm/c
(of the order of the size of nuclei) after the collision and thus provides information
about the early-time thermalization achieved in the collision \cite{Snellings02}. A large elliptic flow
of all charged particles near midrapidity was reported by STAR Collaboration.
The FOPI, INDRA and PLASTIC BALL Collaboration \cite{Luka05, Andro01} are actively involved in measuring the excitation
function of elliptic flow from Fermi energies to relativistic energies. Interestingly, at intermediate energies
($E_{lab}\approx 100 MeV/nucleon$) change from in-plane emission 
(rotation like behaviour) to squeeze-out is predicted
\cite{Soff95, Crochet97} whereas at relativistic energies ($E_{lab}\approx 5 GeV/nucleon$) 
the opposite change from the squeeze-out to in-plane enhancement is observed. 
Elliptic flow requires reinteractions within the produced
matter as a mechanism for transferring the initial spatial deformation of the reaction zone in noncentral collision
onto momentum space. It is thus plausible to expect that the largest elliptic flow signal is produced in
the hydrodynamic limit and an almost linear increase in its value with the particle transverse momentum
below 1.5 GeV/c. In the hybrid model of combining the hydrodynamic model with the RQMD transport model \cite{Sorge95}
and choosing certain effective equation of state, it is possible to obtain an elliptic flow that is comparable to the
measured ones in heavy-ion collisions at both SPS and RHIC energies \cite{Teaney01}.
     The experimental result shows that elliptic flow first increases with particle transverse
momentum and then levels off. The dependence of elliptic flow on both the charged particle multiplicity
\cite{Acker01, Roland02} and the particle pseudorapidity \cite{Roland02} have also been measured.
A complete study of excitation function of transverse momentum and energy dependence of elliptic flow in
the entire energy region can provide useful information about nucleon-nucleon interaction related to nuclear equation
of state. In literature, many attempts have already been made with hard equation of state with free N-N cross-section
and soft EOS with reduced nucleon-nucleon cross-section with and without momentum 
dependent interactions and also tried to explore different aspects of 
directed sideward flow. This study is in
continuation with our previous study \cite{kumar10}, in which we have shown 
that experimental balance
energies can be explained well with reduced isospin dependent 
NN cross-section with hard equationof state. In the present study our aim 
is to pin down the relation between balance energy
and transition energy. Is there any relation between these two energies. Whether there is any
mass dependence or not. For the present study, the isospin dependent quantum 
molecular Dynamics (IQMD) model is used to generate the phase space
of nucleons \cite{Hartnack98}.\\

\section{Results and Discussion}
We study the elliptic flow using a stiff equation of state along with isospin-dependent reduced cross-sections
($\sigma$= 0.9 $\sigma_{NN}$), by simulating various reactions. The time
evolution of the reaction is followed upto 200 fm/c. This is the time at which flow saturates
for lighter as well as for heavier systems. For this study, the reactions of
$^{40}Ar_{18}~+~^{45}Sc_{21}$ ($\hat{b}=0.4$, L=0.5L),
$^{93}Nb_{41}~+~^{93}Nb_{41}$($\hat{b}=0.3$, L=0.7L), 
$^{139}La_{57}~+~^{139}La_{57}$($\hat{b}=0.3$, L=0.8L),
and $^{197}Au_{79}~+~^{197}Au_{79}$(b=2.5fm, L=L)
are simulated, where L is the Gaussian width. As
mentioned in Ref. \cite{Hartnack98}, in IQMD the value of Gaussian width L depends
on the size of the system. For Au nuclei L=8.66 $fm^2$ and for Ca nuclei L=4.33$fm^2$.
$\hat{b}$ is the scaled impact parameter is defined as $\hat{b}=\frac{b}{b_{max}}$ (where b is
particular impact parameter in Fermi(fm) and $b_{max}=1.12(A_{T}^{1/3}+A_{P}^{1/3})$),$A_{T}$ 
and $A_{P}$ is the mass of target and projectile respectively. 
The choice of impact parameter is guided by the
 experimentally extracted information
\cite{Westfall93,Pak96,Cussol02}.  These reaction have been performed at their
corresponding balance energies. The above reactions were simulated between 40 and
1200 MeV/nucleon using the hard equation of state along with isospin-dependent 
reduced cross-sections.
\begin{figure}
\includegraphics[width=0.7\textwidth]{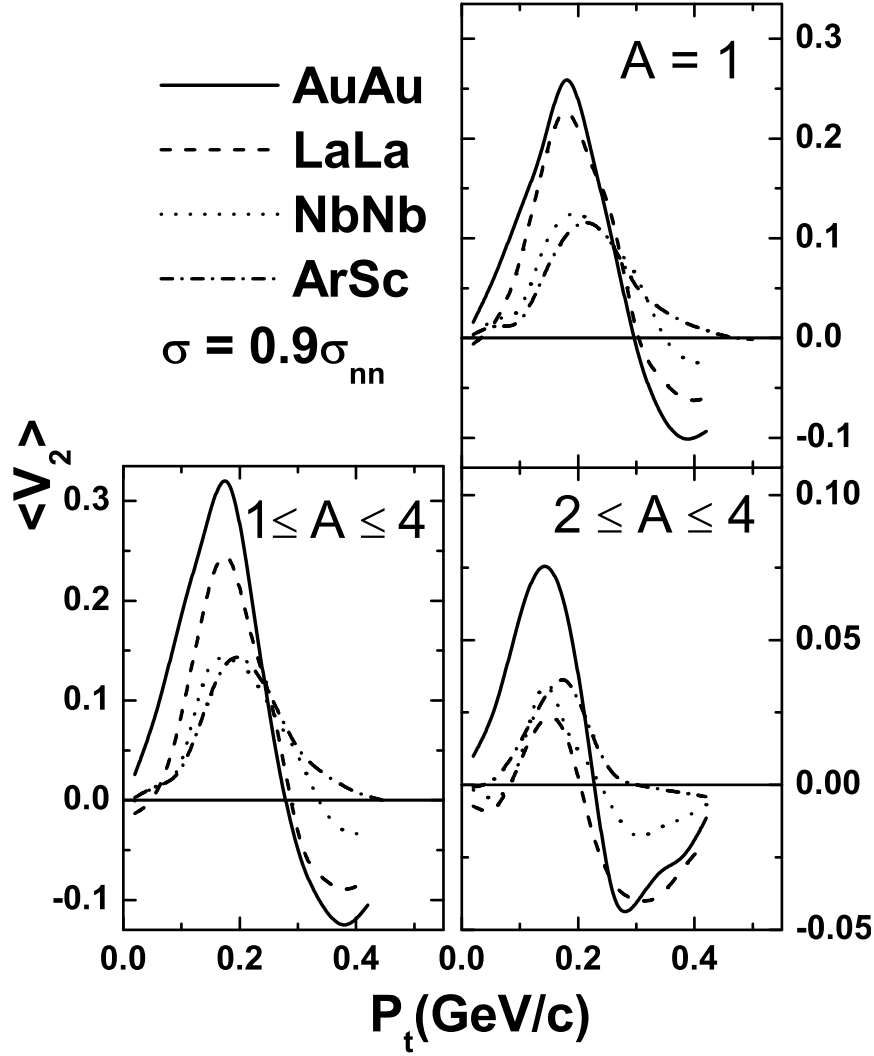}
\caption{\label{fig:1} Transverse momentum dependence of the elliptical 
flow at E = 200 MeV/nucleon. The different
lines in the figure show the variation with diferent system mass and different panels shows the fragments of different mass range.}
\end{figure}                              
\begin{figure}
\includegraphics[width=0.7\textwidth]{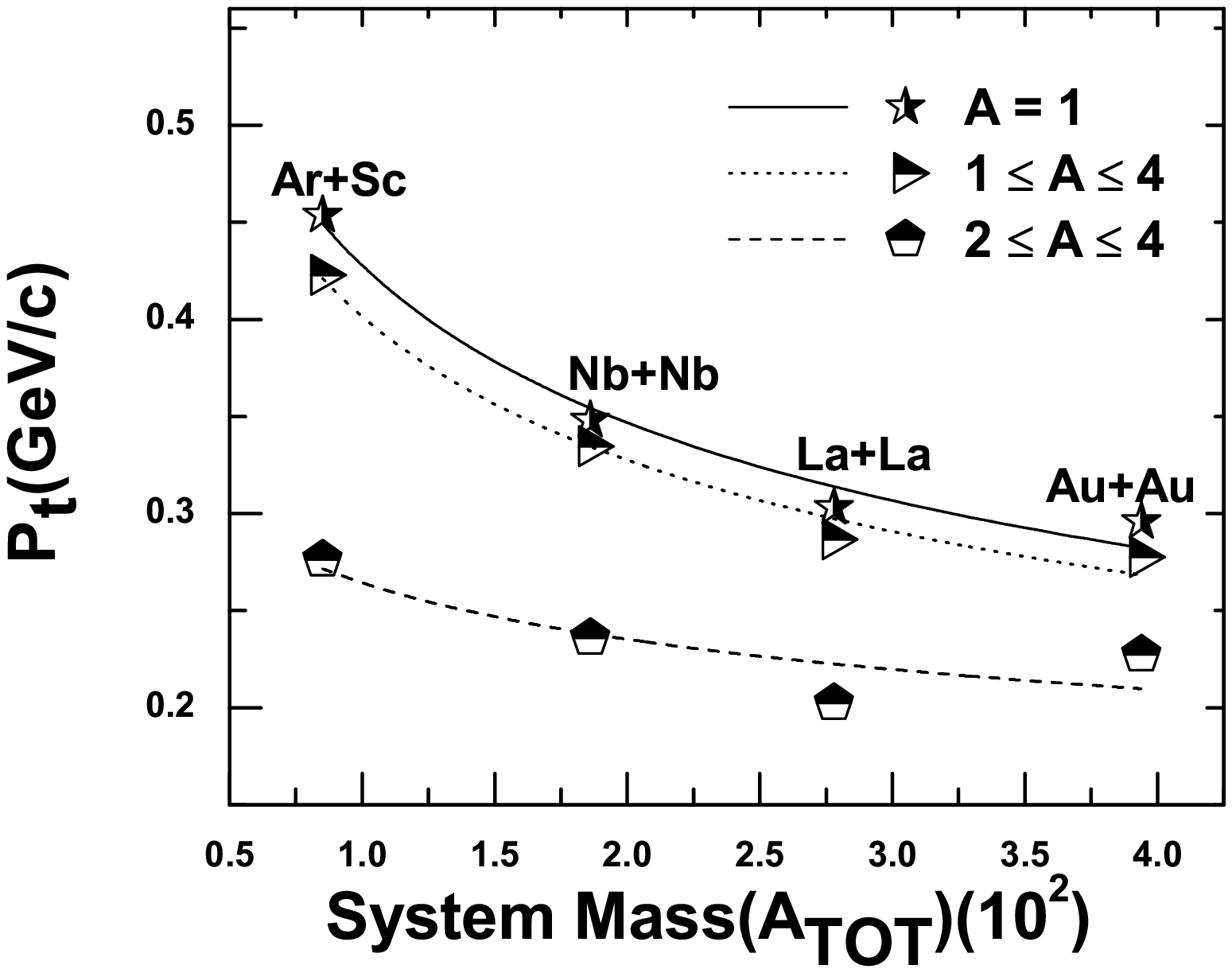}
\caption{\label{fig:2} Transverse momentum dependence ($P_{t}$) for 
elliptic flow as a function of the combined system mass for different 
systems with fragments of different mass range}
\end{figure}
The phase space generated by the IQMD model has been analyzed using
the minimum spanning tree (MST) \cite{Aich91} method. The MST method binds two nucleons
in a fragment if their distance is less than 4 fm. In
recent years, several improvements have also been suggested. One
of the improvements is to also imply momentum cut of the order of Fermi
momentum. This method is dubbed as MSTM method \cite{Sing00a}. The entire calculations are performed
at $t = 200$ fm/c i.e. (Saturation time).\\
\begin{figure}
\includegraphics[width=0.7\textwidth]{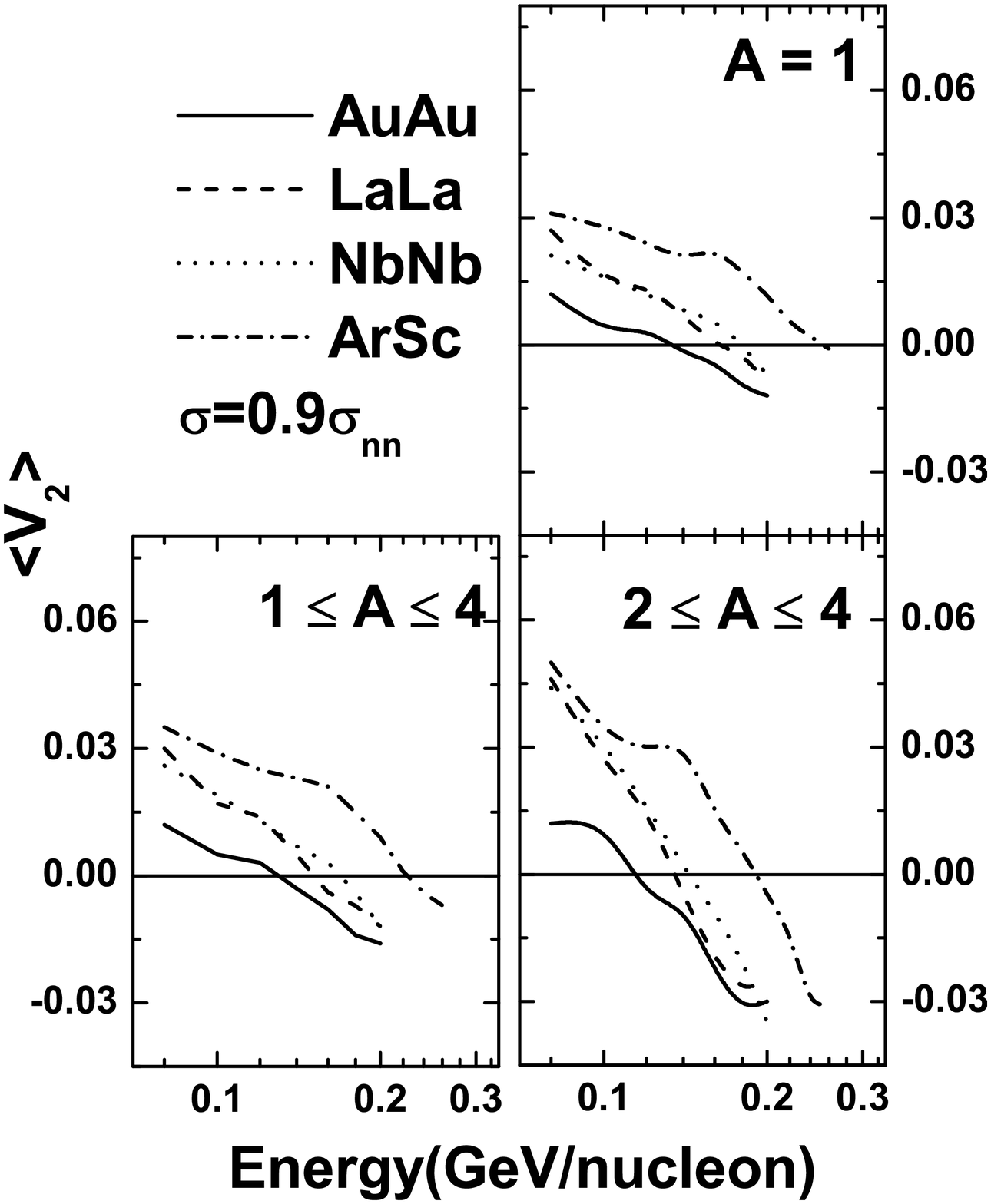}
\caption{\label{fig:3} Variation of the elliptic flow, with beam energy at $|y|$ = $|\frac{y_{c.m}}{y_{beam}}| \le 0.1$
for different reactions.}
\end{figure}
The elliptical flow is defined as the average difference between the square of the x and y components
of the particle's transverse momentum. Mathematically, it can be written as
\begin{equation}
v_2 = \langle\frac{p_x^2 - p_y^2}{p_x^2 + p_y^2}\rangle,
\end{equation}
where $p_x$ and $p_y$ are the x and y components of the momentum. The
$p_x$ is in the reaction plane, while, $p_y$ is
perpendicular to the reaction plane.\\
\begin{figure}
\includegraphics[width=0.7\textwidth]{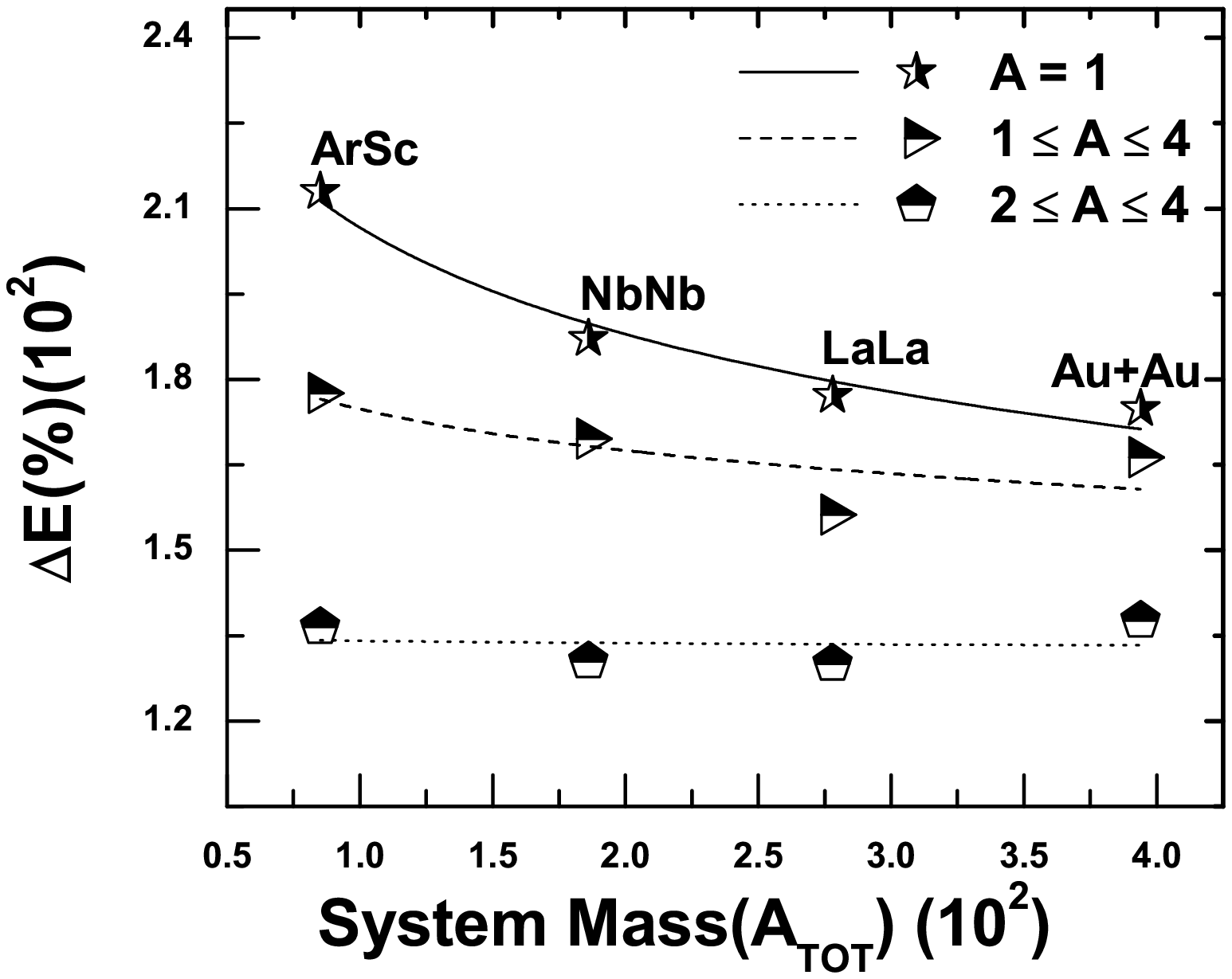}
\caption{\label{fig:4} Difference of transition energy and balance energy as a function of 
mass of system.}
\end{figure}
\begin{figure}
\includegraphics[width=0.7\textwidth]{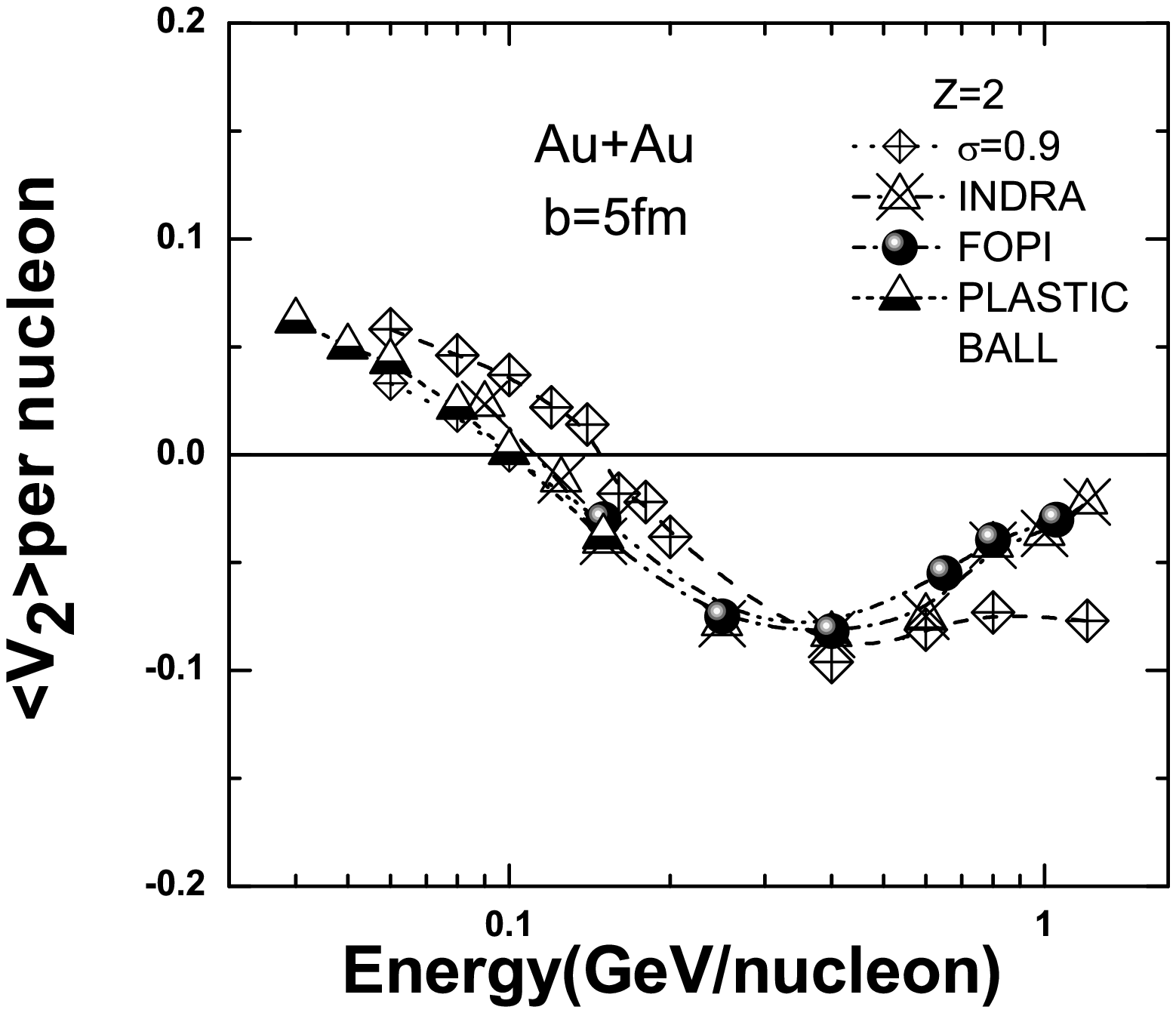}
\caption{\label{fig:5} Energy dependence of the elliptic flow for Au+Au systems 
and its comparison with the experimental data.}
\end{figure}
In the Fig.1, we display the transverse momentum dependence of elliptical flow for the free particles,
light chaged particles. A Gaussian-type behavior is observed in all cases.  Note that this
elliptical flow is integrated over entire rapidity range. It is
also evident from the figure that the peaks of the Gaussian shifts toward lower
values of $P_t$ for heavier fragments.
This is due to the fact that the free and light charged particles feel the mean field
directly, while heavy fragments have weaker sensitivity \cite{Yan07}.
In the Fig.2, we display the system size dependence of the transverse momentum at which  $v_2$
becomes zero for different system and different fragments. The value of transverse momentum decreases
with system mass because in heavier systems there are more Coulomb repulsions
than that of lighter systems. This dependence is again fitted with the power law of the kind
\begin{equation}
E_{Trans} = C(A_{tot}^{-\tau})
\end{equation}
 The value of $\tau$ is found to increase with increase in fragment mass range. As in case of $A=1$
it is $-0.30 \pm 0.03$, for $1 \le A \le 4$ it is $-0.29 \pm 0.02$, for $2 \le A \le 4$ it is
$-0.17 \pm 0.07$.
  In the fig.3, we display the variation of the elliptic flow $v_2$ for free
nucleon, light charge particle (LCP's) over midrapidity region as a 
function of energies. The free particles and LCP's, which
originate from the participant zone, show a systematic behavior with the beam energy and with the
composite mass of the system as well as with the fragments of different mass range. The elliptical flow
for these particles is found to become more negative with the increase in the composite
mass of system and with the increase in the beam energy as well as with the fragments of different mass range.
The heavier the system, the greater the Coulomb repulsion and more negative is the elliptical flow.\\
 The elliptical flow is found to show a transition from in-plane to out-of-plane at a certain beam energy
known as transition energy for mid-rapidity region. This is due to the change in the rotational behavior
into expansion with increase in the incident energy.\\
In the fig.4, we display the system size dependence of the difference of
transition energy($\Delta E(\%)=\frac{E_{t}-E_{b}}{E_{b}}\times$100) extracted from the 
fig.3 for different fragments and
balance energy($\sigma=0.9 \sigma_{NN}$)studied in ref. \cite{kumar10}. For the fragment of
low mass range this difference show slight decrease as we increase the
system mass then other effect comes into play (i.e. expansion of participant and shadowing of spectator matter)
whereas for light charge particle's constant line is obtained which means the additional effect
is independent of system mass. It shows that transition energy and 
balance energy are closer in heavy
mass system as compared to lighter systems. This happen due to increase of neutron,
the number of collision increases and hence leads to decrease in transition energy.
 This dependence is again fitted with the power law.
 The value of $\tau$ is found to increase with increase in fragment mass range. 
As in case of $A=1$ it is  $-0.14 \pm 0.017$, for $1 \le A \le 4$ it is 
$-0.061 \pm 0.036$, for $2 \le A \le 4$ it is$-0.0043 \pm 0.031$.\\
 In the fig.5, we show  $v_2$ at midrapidity $|y|$ = $|\frac{y_{c.m}}{y_{beam}}| \le 0.1$ 
for Z=2 as a function of incident energy. The rapidity cut is in accordance 
with the experimental findings. The theoretical results are compared with the 
experimental data extarcted by  INDRA, FOPI and PLASTIC BALL
collaborations\cite{Luka05,Andro01}. With the increase in the incident energy, elliptical
flow $v_2$ changes from positive to negative values exhibiting a transition from the in-plane to
out-of-plane emission of nucleons.  This is because of the fact that
the mean field, which contributes to the formation of a rotating compound system, becomes less
important and the collective expansion process based on the nucleon-nucleon scattering starts to be predominant.
The maximal negative value of $v_2$ is obtained around $E = 500$
MeV/nucleon with reduced isospin dependent cross-section.
This out-of-plane emission decreases again towards the higher incident energies.
This happens due to faster movement of the spectator matter after
$v_2$ reaches the maximal negative value \cite{Andro01}.
This trend is in agreement with experimental findings. A close agreement with data
is obtained in the presence of hard equation of state and with reduced isospin dependent cross-section
for Z=2 particles. Similar results and trends  have also been reported by Zhang et.al. in their
recent communication \cite{Zhan06}.\\

\section{Conclusion}
By using the IQMD model, we have studied correlation between transition
energy and balance energy. We have investigated the elliptical flow of
fragments for different reacting systems at incident energies between 40 and 1200 MeV/nucleon
using isospin-dependent quantum molecular dynamics (IQMD) model. The elliptical flow is found to show a transition
from in-plane to out-of-plane at a certain beam energy in mid-rapidity region. Our calculation with a stiff equation of
state and reduced isospin dependent nucleon-nucleon cross-section 
($\sigma=0.9\sigma_{NN}$) is in good agreement with the experimental findings.
The difference between balance and transition energy decreases with increase in the composite mass of colliding
nuclei. This tells us that due to increase of neutron to colliding nuclei, the difference
between two energies decreases.\\
 
\begin{acknowledgments}
    Authors are thankful to Prof. J. Aichelin for discussion during his visit to Panjab University
Chandigarh under Indo French joint International project.

\end{acknowledgments}

\end{document}